\documentclass[twocolumn,superscriptaddress,nofootinbib,floatfix,aps,prb,citeautoscript,longbibliography,10pt]{revtex4-2}
\usepackage[utf8]{inputenc}
\usepackage[T1]{fontenc}
\usepackage{lineno}

\usepackage{graphicx}
\usepackage{color}
\usepackage{array}
\usepackage{dcolumn}
\usepackage{bm}
\usepackage{multirow}
\usepackage{chemformula}

\usepackage{booktabs}
\usepackage{tabularx}
\usepackage{comment}
\usepackage{tikz}
\usepackage{xstring}

\usepackage{graphicx}
\usepackage{amsmath}
\usepackage{float}
\usepackage{orcidlink}
\usepackage[capitalise]{cleveref}

\newcommand{\alex}{\textsc{Alexandria}}

\newcommand{\rpm}{\raisebox{.2ex}{$\scriptstyle\pm$}}

\begin{document}

\newcommand{\bochum}{Research Center Future Energy Materials and Systems of the University Alliance Ruhr and ICAMS, Ruhr University Bochum, Universitätsstraße 150, D-44801 Bochum, Germany}
\newcommand{\ETH}{Department of Materials, ETH Zürich, Zürich, CH-8093, Switzerland}

\title{Universal Machine Learning Potentials under Pressure}

\author{Antoine Loew\,\orcidlink{0009-0008-5018-4895}}
\affiliation{\bochum}
\author{Jonathan Schmidt\,\orcidlink{0000-0001-5685-6404}}
\affiliation{\ETH}
\author{Silvana Botti\,\orcidlink{0000-0002-4920-2370}}
\author{Miguel A. L. Marques\,\orcidlink{0000-0003-0170-8222}}
\affiliation{\bochum} 

\date{\today}

\begin{abstract}
Universal machine learning interatomic potentials (uMLIPs) represent arguably the most successful application of machine learning to materials science, demonstrating remarkable performance across diverse applications. However, critical blind spots in their reliability persist. Here, we address one such significant gap by systematically investigating the accuracy of uMLIPs under extreme pressure conditions from 0 to 150~GPa. 
Our benchmark reveals that while these models excel at standard pressure, their predictive accuracy deteriorates considerably as pressure increases. This decline in performance originates from fundamental limitations in the training data rather than in algorithmic constraints. In fact, we show that through targeted fine-tuning on high-pressure configurations, the robustness of the models can be easily increased. These findings underscore the importance of identifying and addressing overlooked regimes in the development of the next generation of truly universal interatomic potentials.
\end{abstract}

\maketitle

\section{Introduction}

High-pressure materials science has emerged as a promising pathway for discovering new materials with valuable applications, ranging from superconductivity to nanostructuring~\cite{McMillan2002,Zhang2017,McMahon2006}. At elevated pressures, materials undergo dramatic structural, electronic, and reactive changes, such as phase transitions, metallization, polymerization, and the formation of exotic compounds~\cite{Loubeyre2020,Moog2021}. These phenomena are not only central to understanding planetary interiors, where pressures exceed hundreds of gigapascals, but also critical for synthesizing advanced materials with tailored properties, such as ultra-hard nano-materials or energy-dense polymorphs~\cite{Grochala2007,Popov2002,Eremets2004,Tang2023}. For instance, the metallization of hydrogen under extreme compression~\cite{Eremets2011}, the discovery of non-molecular CO$_{2}$ phases~\cite{Iota1999,Giordano2007}, and the pressure-induced polymerization of organic molecules~\cite{Li2021} exemplify how high-pressure environments enable the creation of novel states of matter with groundbreaking applications.

Traditional computational approaches to investigating high-pressure chemistry have relied extensively on quantum mechanical methods, particularly density functional theory (DFT), which have provided invaluable fundamental insights into pressure-induced phenomena~\cite{Liu2009,Sun2005}. Recent comprehensive reviews highlight the extensive application of \textit{ab initio} simulations across the gigapascal pressure regime, demonstrating their crucial role in predicting and understanding high-pressure behavior~\cite{Napirkowska2023, Zeller2024, Rychkov2020}. However, these quantum mechanical approaches suffer from computational costs that scale prohibitively with both system size and simulation timescales, strongly limiting their utility for exploring complex multi-component systems, large-scale structural transformations, or kinetically controlled processes under dynamic pressure conditions.

The classical limitations in computational cost associated with \textit{ab initio} methods have undergone a paradigmatic shift with the emergence of machine learning interatomic potentials (MLIPs),  which bridge the accuracy of quantum mechanical approaches with the efficiency of classical force fields~\cite{behler_perspective_2016,Thompson2015,graser_machine_2018,schmidt_recent_2019,Unke2021}. With the rapid advancement of MLIPs, various computational frameworks were developed to integrate these methods~\cite{schutt2019schnetpack,Wang2018}, while an increasing number of \textit{ab initio} databases have begun to offer extensive datasets of high-quality and consistent DFT calculations~\cite{Jain2013,Kirklin2015,Curtarolo2012,Schmidt2024_1,Scheidgen2023}. This combination of software and data led to the development of universal MLIPs (uMLIPs), i.e. foundation models that are capable, without further training, of describing a diverse materials space covering the whole periodic table. One of the first examples was M3GNet~\cite{chen_universal_2022}, an extension of MEGNet that incorporated three-body interactions within its graph neural network architecture, thereby enriching the local chemical environment representation provided to the model. Following this development, numerous uMLIPs have been proposed~\cite{Neumann2024, Park2024, Liao_2023, Deng2023, Choudhary2021, mattersim}, trained on increasingly larger and more diverse datasets, comprising up to millions of compounds and tens to hundreds of millions of DFT calculations~\cite{Barroso2024, Schmidt2024_1,Deng2023, Peng2025}, significantly broadening the applicability and robustness of ML-driven materials materials science.

Despite the impressive scale and diversity of existing datasets, many universal MLIPs still fail to generalize to specific regimes of interest, such as high pressures. In such cases, the lack of targeted training data can lead to significant inaccuracies, limiting the applicability of these models. In this context, we argue that the development and validation of uMLIPs specifically adapted to high-pressure environments is crucial. Such models have the potential to bridge the gap between quantum mechanical accuracy and computational efficiency, enabling reliable predictions in pressure regimes that remain challenging for conventional approaches. In this work, we tackle this problem by introducing a new dataset comprising 190 thousand compounds with a total of 32 million atomic single-point calculations under varying pressure. Using this dataset, we demonstrate the limitations of state-of-the-art uMLIPs in this regime and explore strategies for fine-tuning these models to restore their predictive capability under high-pressure conditions.

\section{Results}
\subsection{Dataset of calculations under pressure}

The database of \textit{ab initio} calculations under pressure used in this work is summarized in \cref{table:pressuredatabase}. Our dataset extends the existing \alex\ database~\cite{Schmidt2024_1}, which has been used to train several state-of-the-art uMLIPs~\cite{Neumann2024} and develop the largest training datasets freely available at the moment~\cite{mattersim, Barroso2024}. 

The 0~GPa dataset consists of 190 thousand distinct structures covering the complete periodic table. Each of these structures was then re-evaluated using DFT with the Perdew-Burke-Ernzerhof (PBE) exchange correlation functional~\cite{Perdew1996}, consistently with the calculation setup used for the training data, across a predefined set of pressure values ranging from 0~GPa to 150~GPa. For each pressure value, full ionic relaxations were performed to obtain equilibrium volumes, atomic positions, and total energies. The resulting high-pressure dataset comprises a total of 32 million single-point DFT calculations. All calculations were performed using the same exchange-correlation functional and computational parameters as in the original \alex\ dataset to ensure consistency. The final dataset (see \cref{table:pressuredatabase}) includes, for each pressure and material, the relaxed crystal structure, total energy, atomic forces and stress tensors, together with the atomic configurations along the relaxation path.

\begin{table}[htp]
\centering
\caption{Summary of the pressure dataset develop in this work. The missing materials at ambient pressure are due to older calculations for which we do not have the geometry optimization paths in \alex. The missing materials under pressure concern calculations that did not converge.}
\label{table:pressuredatabase}
\begin{tabular*}{\columnwidth}{@{\extracolsep{\fill}} l c c c c c c c c}
\toprule
Pressure (GPa)            & 0    & 25    & 50     & 75   & 100   & 125   & 150  \\
\midrule
Materials [x10$^3$]               & 162  & 188  & 190  & 188  & 188  & 188  & 187 \\
Atomic configurations [x10$^6$]    & 3.2  & 4.0  & 6.0  & 4.0  & 4.7  & 5.1  & 5.2 \\

\bottomrule
\end{tabular*}
\end{table}

\begin{figure}[htb!]
  \centering
  \includegraphics[width=0.9\columnwidth]{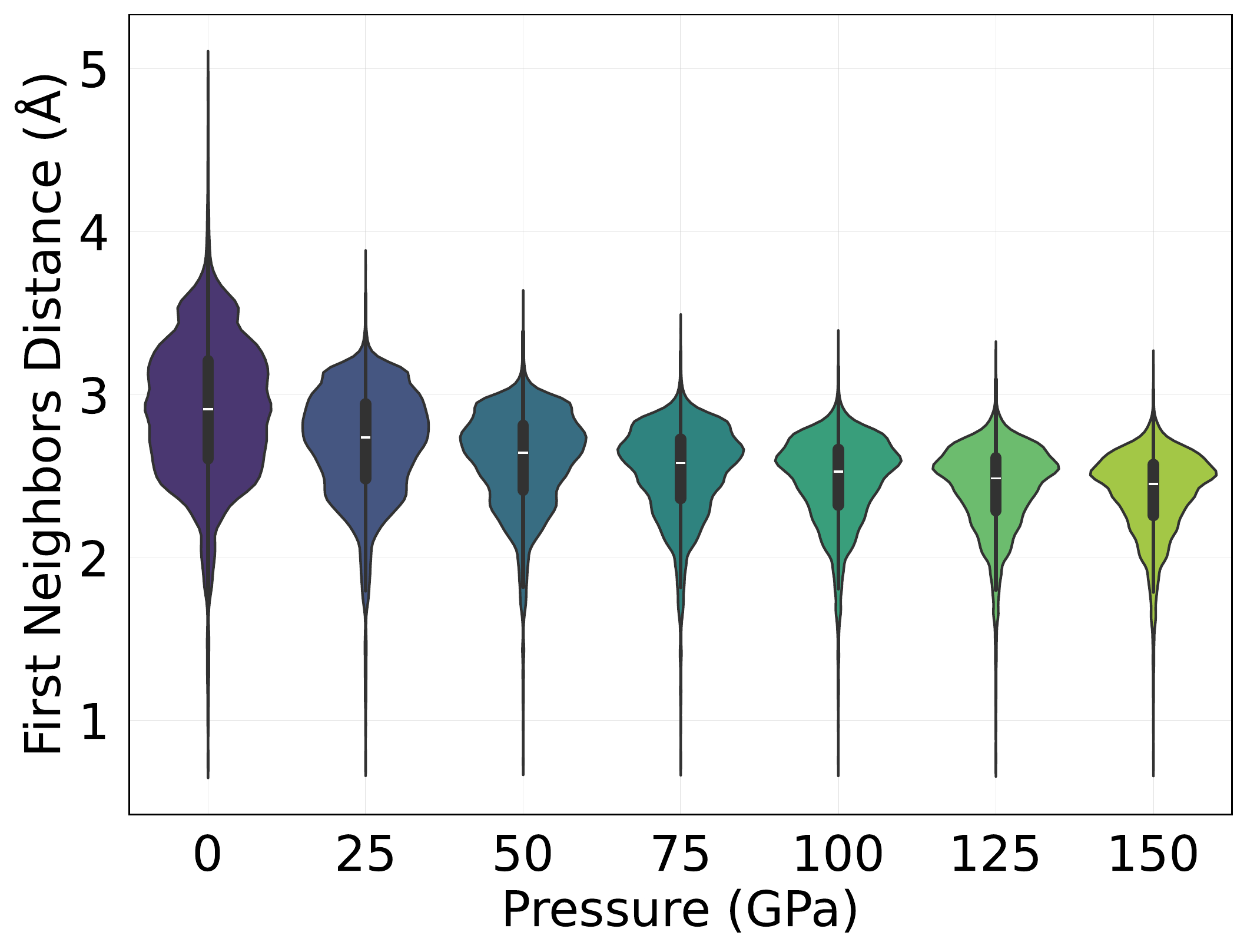}\\
  \caption{Violin plot showing the distribution of first neighbor distances in the DFT datasets across the pressure range.}
  \label{fig:violin_neigh}
\end{figure}

As illustrated in \cref{fig:violin_neigh}, the distribution of first-neighbor distances in the dataset systematically narrows and shifts toward shorter values with increasing pressure, in line with expectations. We observe a systematic decrease in its maximum value, going from nearly 5~\AA{} at ambient pressure to approximately 3.3~\AA{} at the highest pressure investigated. Contrary to the overall contraction of atomic environments, the minimum first neighbor distance remains relatively stable with a slightly decrease from 0.74~\AA{} at ambient pressure to 0.72~\AA{} at 150~GPa. Obviously, these distances correspond to the very strong, and rather uncompressible, covalent bonds between first row chemical elements.

\begin{figure}[!htb]
  \centering
  \includegraphics[width=1\columnwidth]{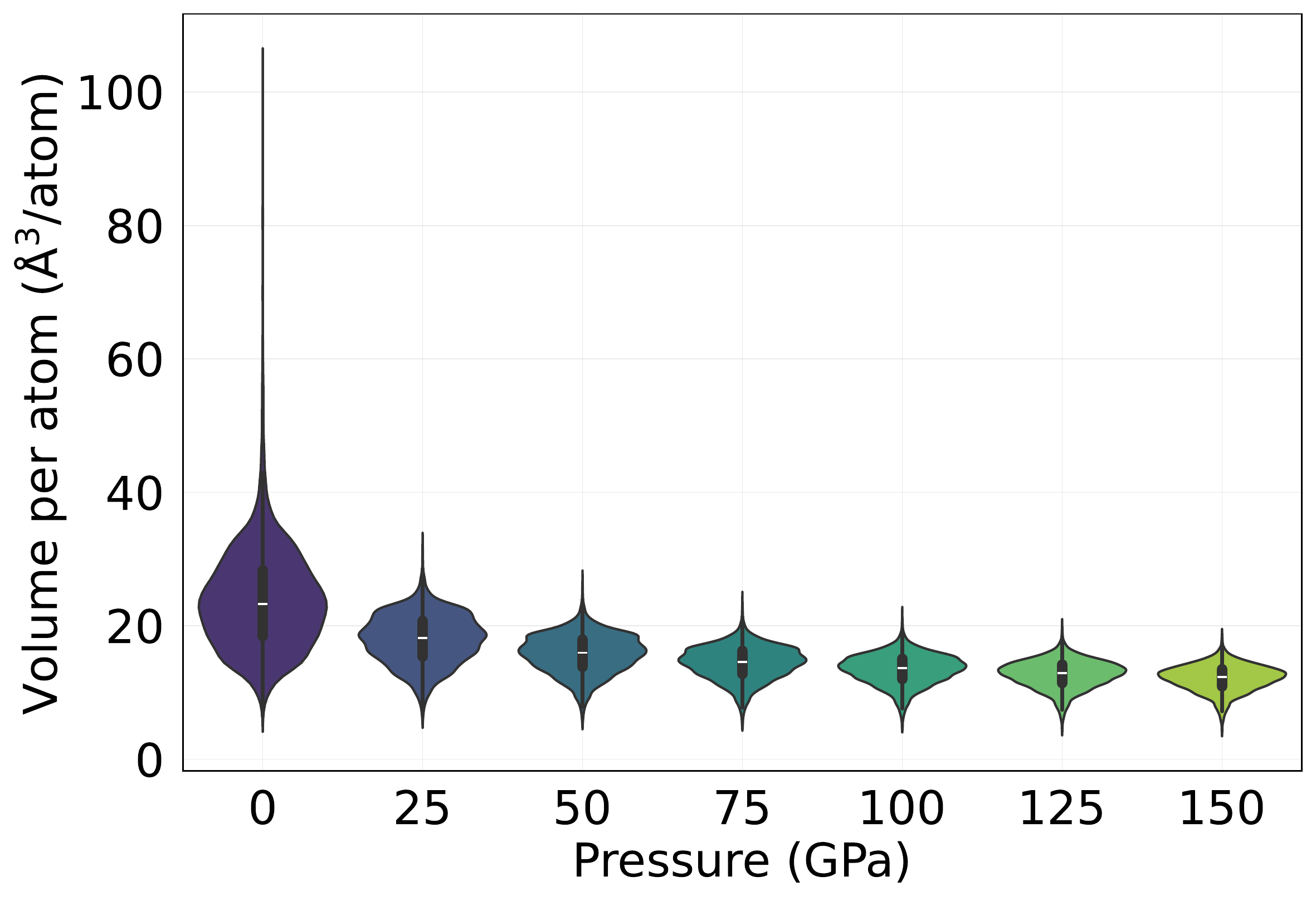}\\
  \caption{Violin plot showing the distribution of unit cell volumes per atom in the DFT datasets across the pressure range.}
  \label{fig:volume_pressure_violin}
\end{figure}

\cref{fig:volume_pressure_violin} shows a  comparable pressure-dependent trend for the volume per atom. At ambient pressure, the distribution of volumes per atom is broad, ranging from approximately 10 to 40~\AA$^3$/atom, with a long tail extending beyond 100~\AA$^3$/~atom. The tail corresponds to layered compounds whose volume is overestimated due to the lack of the van der Waals interaction in the PBE approximation used. Upon applying pressure, the distribution narrows significantly, and is shifted toward lower volumes per atom, reaching roughly 20~\AA$^3$/atom at the highest pressure. The distribution becomes increasingly uniform under compression, reflecting extensive structural changes within the datasets with increasing pressure.

Although this trend is a natural consequence of compression under high-pressure conditions, the accompanying shift in atomic environments constitutes a fundamental structural change that cannot be captured in conventional DFT datasets generated at ambient pressure.

\subsection{Performance and fine-tuning of universal MLIPs}

We employ the high-pressure datasets both to benchmark the generalization capabilities of the best-performing uMLIPs and to fine tune these models to improve their prediction of structural properties under pressure. To ensure a fair and representative comparison between fine-tuned and non-fine-tuned models, we partitioned the datasets at the level of the 190 thousand materials using a 90\%–5\%–5\% split for training, validation, and test sets, respectively. This approach prevents data leakage across subsets by assigning all frames from a given relaxation trajectory to the same partition, ensuring that no structural information from a single pressure path appears in multiple splits. The specific criteria used for data selection, as well as the thresholds applied to energy, force, and stress during subsampling, are detailed in \cref{sec:method}. Note that there can be a small amount of data leakage at 0~GPa as the \alex\ database or systems from the \alex\ database were used to pretrain some of the universal force fields. 

The uMLIPs selected for benchmarking were chosen to represent the diversity of available atomic foundation models as well as their historical evolution. For consistency, and when possible, we adopt the model naming used in the Matbench benchmarking~\cite{Riebesell_matbench}.
M3GNet~\cite{chen_universal_2022} and MACE-MPA-0~\cite{batatia_mace_2023} represent two of the earliest foundation models. M3GNet was trained on the Materials Project database which, while widely used, offers a comparatively limited dataset relative to more recent uMLIPs. In contrast, MACE-MPA-0 represents an updated version of the original MACE architecture, trained on MPtrj~\cite{Deng2023} and subsampled \alex~\cite{Barroso2024}, referred to as the MPA dataset. MACE-MPA-0 uses density renormalization to improve performance for systems under high pressure.
SevenNet-MF-OMPA~\cite{Park2024}, based on the NequIP framework~\cite{batzner_e3-equivariant_2022}, is a highly data-efficient model trained on OMAT~\cite{Barroso2024}-MPtrj-subsampled \alex\ (OAM) dataset. DPA3-v1- ~\cite{Zhang2025} represents a large-scale atomic model designed for broad applicability, having been trained on 163 million structures collected from the literature.
GRACE-2L-OAM~\cite{Bochkarev_2024} builds on the Atomic Cluster Expansion (ACE) formalism~\cite{Drautz2019} with a graph-based architecture, trained on the OAM dataset.
ORB-v3-Conservative-Inf~\cite{rhodesOrbv3AtomisticSimulation2025} represents the third generation of the ORB models~\cite{Neumann2024}, incorporating conservative forces and removing any limitation on the number of neighbors per atom.
MatterSim-v1~\cite{mattersim} is a modern iteration of M3GNet trained on a proprietary dataset of 17 million structures.
Lastly, eSEN-30M-OAM~\cite{esen}, one of the most recent models, employs multiple techniques to ensure a smooth potential energy surface (PES) and is trained on the OAM dataset.

To assess the potential benefits of fine-tuning for improving model performance under high-pressure conditions, we selected two representative uMLIP models for refinement: MatterSim-ap-ft-0 and eSEN-ap-ft-0. These models were fine-tuned using our high-pressure dataset to evaluate how incorporating pressure-specific atomic environments influences their predictive accuracy and generalization capabilities.

\begin{figure}[htb!]
  \centering
  \includegraphics[width=1\columnwidth]{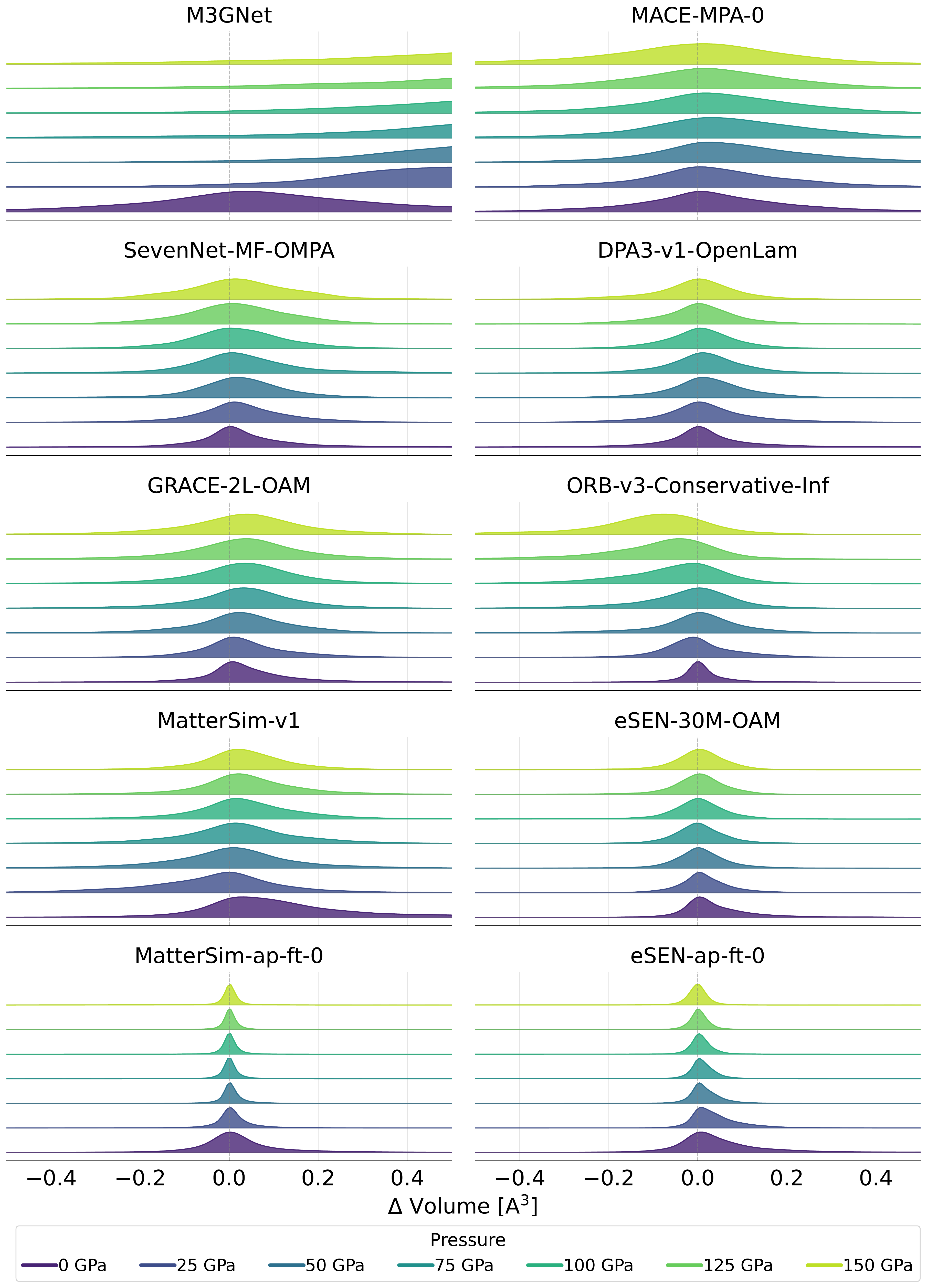}\\
  \caption{Volume error after structural optimization: comparison between uMLIP predictions and PBE reference data.}
  \label{fig:volume_error_pressure_violin}
\end{figure}

\begin{table}[htp]
\centering
\caption{Mean absolute error of the uMLIP volume predictions (in \AA$^3$)  with respect to the PBE reference values across the range of pressures P$_{x}$ ($x$ in GPa)}
\label{tab:volume_mae_pressure}
\begin{tabular}{lccccccc}
\hline
Model & P$_{0}$ & P$_{25}$ & P$_{50}$ & P$_{75}$ & P$_{100}$ & P$_{125}$ & P$_{150}$ \\
\hline
M3GNet & 0.42 & 1.28 & 1.56 & 1.58 & 1.50 & 1.44 & 1.39 \\
MACE-MPA-0 & 0.22 & 0.16 & 0.18 & 0.19 & 0.19 & 0.20 & 0.22 \\
SevenNet-MF-OMPA & 0.11 & 0.10 & 0.10 & 0.11 & 0.10 & 0.11 & 0.11 \\
DPA3-v1-OpenLam & 0.11 & 0.09 & 0.09 & 0.08 & 0.08 & 0.08 & 0.08 \\
GRACE-2L-OAM & 0.12 & 0.10 & 0.11 & 0.12 & 0.12 & 0.13 & 0.14 \\
ORB-v3-Conservative-Inf & \textbf{0.05} & 0.08 & 0.10 & 0.11 & 0.15 & 0.18 & 0.22 \\
MatterSim-v1 & 0.23 & 0.14 & 0.13 & 0.13 & 0.11 & 0.11 & 0.10 \\
eSEN-30M-OAM & 0.09 & 0.05 & 0.06 & 0.06 & 0.06 & 0.05 & 0.06 \\
MatterSim-ap-ft-0 & 0.21 & \textbf{0.04} & \textbf{0.02} & \textbf{0.02} & \textbf{0.02} & \textbf{0.02} & \textbf{0.02} \\
eSEN-ap-ft-0 & 0.19 & 0.06 & 0.03 & 0.03 & \textbf{0.02} & \textbf{0.02} & \textbf{0.02} \\
\hline
\end{tabular}
\end{table}

\begin{figure}[htb!]
  \centering
  \includegraphics[width=1\columnwidth]{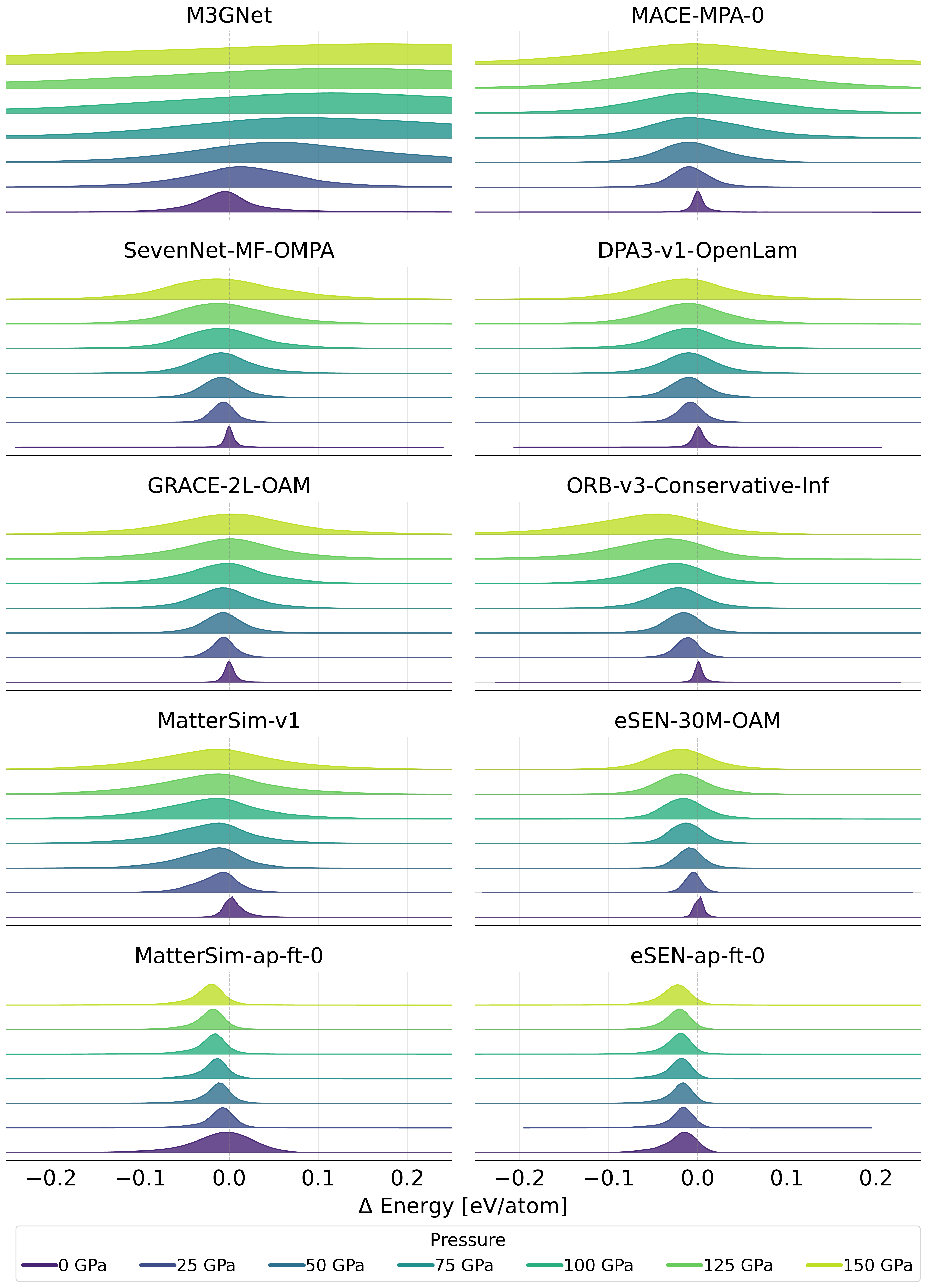}\\
  \caption{Error in predicted energy/atom after structural optimization: comparison between uMLIP predictions and PBE reference data.}
  \label{fig:energy_pressure_violin}
\end{figure}

The performance of each uMLIP model, along with the fine-tuned versions, were evaluated on the test set across the entire pressure range. Results are reported in \cref{fig:volume_error_pressure_violin} and \cref{fig:energy_pressure_violin} and in \cref{tab:volume_mae_pressure} and \cref{tab:energy_mae_pressure}. Errors are calculated as differences between energies and geometries at the end of the DFT and the uMLIP relaxations as this is, in our opinion, the metric that is most relevant for modern workflows of materials discovery.

A common trend across all models is the systematic deterioration in prediction accuracy as pressure increases, spanning the range from ambient conditions up to 150~GPa. This behavior is particularly visible in the original, untuned models, where both the energy and volume errors exhibit significant increases with pressure, indicating a growing deviation from the reference values.

As shown in \cref{fig:volume_error_pressure_violin} and \cref{tab:volume_mae_pressure}, the volume prediction errors are notably high for the two earliest uMLIP models. M3GNet exhibits consistently larger volume under pressure, with deviations spanning several~\AA$^3$. This behavior is expected, as the Materials Project database, which serves as M3GNet's primary training source, includes relatively few configurations far away from dynamical equilibrium at ambient pressure. As a result, the model tends to underestimate compressibility, leading to a systematic overprediction of equilibrium volumes under pressure. In contrast, the MACE model yields a zero-centered normal distribution of volume errors across the entire pressure range. However, it exhibits a broad spread in the error distribution, even at ambient pressure, ranging approximately from –0.2~\AA$^3$ to 0.4~\AA$^3$. This suggests a lack of accuracy in volume predictions, even under standard thermodynamic conditions.

SevenNet-MF-OMPA and GRACE-2L-OAM exhibit similar performance, with a well-centered and narrow distribution of volume errors at ambient pressure demonstrating the benefits of the one order of magnitude larger training datasets. However, as pressure increases, the error distributions broaden, resulting in a spread of approximately –0.2~\AA$^3$ to 0.2~\AA$^3$ at 150~GPa. 
Orb-v3-Conservative-Inf stands out at ambient pressure, exhibiting an exceptionally narrow error distribution concentrated between –0.05~\AA$^3$ and 0.05~\AA$^3$, indicating highly accurate volume predictions under standard conditions. However, as pressure increases, its performance degrades more rapidly than that of other uMLIPs, systematically overestimating compression. This over pressurization behavior is a distinctive trend not observed in the other models. One explanation could be that, unlike the other models, ORB-v3 removed all rattled or volume-scaled systems from the Omat24 dataset, since these structures led to unphysical behavior in diatomic systems~\cite{rhodesOrbv3AtomisticSimulation2025}. This appears to have improved its equilibrium capabilities, but at the cost of decreased performance at high-pressure.
MatterSim-v1 and DPA3-V1-OPENLAM exhibits a surprising trend, showing improved accuracy under higher pressure compared to ambient conditions. The error distribution of MatterSim-v1 at low pressure spans from –0.15~\AA$^3$ to 0.2~\AA$^3$ with a range does not broaden with increasing pressure. On the contrary, the spread of volume errors tends to contract, indicating enhanced predictive stability under compression. Again we can hypothesize that this due to the strategy of data generation of MatterSim-v1, that included molecular dynamics simulations under pressures reaching up to 1~TPa.
DPA3-V1-OPENLAM performs better than MatterSim-v1 at all pressures and is the second best model in our benchmark.
Finally, eSEN-30M-OAM demonstrates consistently well-behaved volume predictions across the entire pressure range, maintaining a narrow and symmetric error distribution between –0.1~\AA$^3$ and 0.1~\AA$^3$. As it is the only model trained using denoising of non-equilibrium structures~\cite{liao2024generalizing} it would be interesting to explore if the model architecture or training strategy are responsible for its performance.

The fine-tuned versions of MatterSim-v1 and eSEN-30M-OAM show an improvement in volume prediction across the entire pressure range. Although the eSEN-ap-ft-0 model shows a slight degradation in performance at ambient pressure, it achieves a significantly narrower error distribution at 150~GPa, between –0.05\AA$^3$ and 0.05~\AA$^3$. MatterSim-ap-ft-0 demonstrates even greater fine-tuning effectiveness, consistently outperforming eSEN-ap-ft-0 with a sharper and more stable error distribution across all pressure conditions.

\begin{table}[htp]
\centering
\caption{Mean absolute error for energy per atoms predictions (in~meV/atom) across all range of pressure P$_{x}$ (x in GPa)}
\label{tab:energy_mae_pressure}
\begin{tabular}{lccccccc}
\hline
Model & P$_{0}$ & P$_{25}$ & P$_{50}$ & P$_{75}$ & P$_{100}$ & P$_{125}$ & P$_{150}$ \\
\hline
M3GNet & 33.8 & 61.4 & 116.6 & 191.9 & 253.9 & 302.4 & 346.8 \\
MACE-MPA-0 & 6.9 & 22.8 & 41.1 & 61.3 & 84.2 & 107.6 & 163.5 \\
SevenNet-MF-OMPA & 5.0 & 13.6 & 23.7 & 33.2 & 42.8 & 52.1 & 63.7 \\
DPA3-v1-OpenLam & 6.5 & 15.3 & 23.4 & 31.0 & 38.3 & 46.7 & 56.0 \\
GRACE-2L-OAM & 5.8 & 14.3 & 23.6 & 34.7 & 51.6 & 110.9 & 163.3 \\
ORB-v3-Con-Inf & 4.7 & 16.6 & 26.2 & 39.9 & 58.8 & 82.2 & 113.8 \\
MatterSim-v1 & 11.1 & 26.2 & 40.7 & 51.9 & 60.7 & 67.9 & 76.6 \\
eSEN-30M-OAM & \textbf{4.1} & \textbf{10.2} & \textbf{16.5} & \textbf{22.6} & 29.2 & 34.1 & 41.7 \\
MatterSim-ap-ft-0 & 73.5 & 18.3 & 21.0 & 23.3 & 26.1 & 28.4 & \textbf{31.1} \\
eSEN-ap-ft-0 & 24.6 & 22.2 & 23.4 & 24.8 & \textbf{25.8} & \textbf{27.2} & 32.4 \\
\hline
\end{tabular}
\end{table}

We report the error in the energy predicted by each uMLIPs in \cref{fig:energy_pressure_violin} and \cref{tab:energy_mae_pressure}. M3GNet continues to exhibit poor behavior as pressure increases, with large deviations that worsen under compression. All other uMLIPs demonstrate a remarkably narrow normal distribution of energy errors at ambient pressure, typically within \rpm 0.03~eV/atom, consistent with recent benchmarking standards. However, as pressure increases, their energy predictions exhibit significant broadening and a systematic underestimation. Among all models, eSEN-30M-OAM delivers the most accurate energy predictions under pressure, with an error range from –0.08~eV/atom to +0.05~eV/atom at 150~GPa.

In contrast to the volume metric, the fine-tuned models exhibit slightly degraded energy predictions at ambient pressure. This outcome is expected, as with the basic fine-tuning we used there will be some extent of catastrophic forgetting from ambient-pressure data. However, this trade-off results in significantly improved energy predictions across the full pressure range, demonstrating the benefit of targeted adaptation for high-pressure regimes. Using replay strategies during finetuning should reduce this ambient pressure deterioration. In general, the fine-tuned models represent more of a lower bound on the performance achievable with the dataset, and improved results can likely be obtained through hyperparameter optimization or by training the more expensive eSEN model for additional epochs.

\section{Discussion}
\label{sec:conclusion}

In this work, we introduced a large-scale ab-initio dataset containing 30 million DFT calculations under a wide range of pressures, carefully generated using consistent parameters across all configurations. The dataset enabled the benchmarking of several state-of-the-art uMLIPs, as well as fine-tuning selected models. 
Our analysis revealed systematic deviations in performance across the pressure range, with many uMLIPs showing a marked degradation in accuracy when transitioning from ambient pressure to high-pressure regimes. These findings highlight the importance of explicitly including off-equilibrium and high-pressure structures in training datasets to improve the transferability of future uMLIPs. Moreover, our results indicate that different architectural families of uMLIPs exhibit varying extrapolation behaviors. In particular, eSEN-30M-OAM, despite being trained without explicit high-pressure data, demonstrates strong predictive capabilities at elevated pressures, suggesting either better architectural robustness or benefits of denoising of non-equilibrium structures~\cite{liao2024generalizing}.

The dataset introduced here provides researchers with both a means to directly improve uMLIPs and a benchmark for assessing progress in their generalization capabilities. It is important to emphasize that the performance ranking presented in this study should not be regarded as fixed. As more comprehensive datasets are assembled and new architectures are developed, the relative strengths of different models will naturally evolve. Nevertheless, the present results highlight that architecture design and training strategies exert a significant influence on out-of-distribution accuracy and model size factors that should be considered carefully when selecting a model for a specific application. 

Ultimately, this study demonstrates the remarkable extrapolation capabilities of current uMLIPs while also underscoring the considerable potential for further advancement. Achieving this progress will require a coordinated community effort to collect and disseminate large, diverse, and high-quality datasets, ideally with ab-initio methods going beyond the dataset proposed in this work. Such resources will be indispensable for training the next generation of uMLIPs, which should ideally provide robust and reliable predictions across the widest possible range of materials science applications, including systems far from equilibrium at ambient conditions.

\section{Methods}
\label{sec:method}
\subsection{\textit{Ab initio} dataset}
The extension of the \alex\ dataset to higher pressure \textit{ab initio} calculation have been done using the code \textsc{vasp}~\cite{Kresse1996_1,Kresse1996_2}. We used the Perdew-Burke-Ernzerhof (PBE) exchange correlation functional~\cite{Perdew1996} in the same conditions as in the \alex\ dataset~\cite{Schmidt2024_1}.

\subsection{Fine tuning}
\textbf{Fine-tuning Dataset} - The fine-tuning dataset was split into training, validation, and test sets using a standard 90\%–5\%–5\% ratio. All trajectories from a given optimization path for a structure at any pressure were kept within a single subset. To select which points to include from the optimization paths, we first merged all optimization paths of a given structure across all pressures, and then selected all structures that differed by more than 10~meV/atom, starting from the lowest-energy configurations. This resulted in 8,190,860 data points in the training set, 456,358 in the validation set, and 457,745 in the test set. An additional outlier removal step was applied, retaining only structures with energies between –18.0~eV/atom and 10~eV/atom, maximum force components below 100~eV/\AA, and minimum stress components above –200~eV/\AA$^3$. This filtering removed only 0.3\% of the data, keeping 99.7\% of the dataset.

\textbf{eSEN} - The eSEN models were finetuned using the fairchem library~\cite{fairchem} version 1.3.0. We used the AdamW optimizer with a weight decay of 0.001 and an initial learning rate of 0.0004. Learning rate scheduling was handled by a cosine annealing scheduler with 100 total epochs, a minimum learning rate factor of 0.1, and a warmup phase of 5 epochs with a warmup factor of 0.2. Training was conducted with a batch size of 256 divided onto 16 GH200 processors. Gradient clipping was applied with a maximum norm of 100, and an exponential moving average (EMA) of model weights was maintained with a decay rate of 0.999. The L1-norm loss function used weights 20, 200, 200 respectively for energy/atom, forces and stress. 

\textbf{Mattersim} - The MatterSim-v1.0.0-5M model was finetuned using the mattersim repository version 1.1.1. The code was adapted to also use lmdb databases to enable the training on large datasets. We used the AdamW optimizer with an initial learning rate of 2e-4, and step-wise learning-rate reduction every 5 epochs by a factor of 0.95. We trained with a batch size of 256 split onto 8 GH200 processors. Gradient clipping was applied with a maximum norm of 1. The energy-force-stress loss  ratio was adjusted during the training from an initial 1-100-0.1 to 1-1-0.5 and 1-10-1 at 20 and 61 epochs.

\section{Data availability}
\label{sec:data_ava}
The pressure dataset will be available upon acceptance in \alex\  which can be accessed and/or downloaded from \url{https://alexandria.icams.rub.de/} under the terms of the Creative Commons Attribution 4.0 License.  

\section{Code availability}
\label{sec:code_ava}
All code and models developed in this work will be freely available upon acceptance at  \url{https://github.com/hyllios/utils/tree/main/}.

\section{Acknowledgements}

A.L., S.B., and M.A.L.M. acknowledge funding from the Horizon Europe MSCA Doctoral network grant n.101073486, EUSpecLab, funded by the European Union. M.A.L.M. was supported by a collaboration between the Kavli Foundation, Klaus Tschira Stiftung, and Kevin Wells, as part of the SuperC collaboration, and by the Simons Foundation through the Collaboration on New Frontiers in Superconductivity (Grant No. SFI-MPS-NFS-00006741-10). We would like to thank the NHR Centre PC2 for providing computing time on the Noctua 2 supercomputers. J.S. was supported by the European Research Council (ERC) under the European Union’s Horizon 2020 research and innovation program project HERO Grant Agreement No. 810451. Computational resources were provided by ETH Zurich and by the Swiss National Supercomputing Center (CSCS) under project ids s1128 and s1273.
\bibliography{main.bib}

\end{document}